\documentclass[prl,preprint]{revtex4}

\usepackage[dvips]{graphicx}
\usepackage{amsmath}

\begin{document}

\title{Unsteady Crack Motion and Branching in a Phase-Field Model of
  Brittle Fracture}

\author{Alain Karma} \author{Alexander E.~Lobkovsky}
\affiliation{Department of Physics and Center for Interdisciplinary
  Research on Complex Systems, Northeastern University, Boston, MA
  02115}

\begin{abstract}
  Crack propagation is studied numerically using a continuum
  phase-field approach to mode III brittle fracture.  The results shed
  light on the physics that controls the speed of accelerating cracks
  and the characteristic branching instability at a fraction of the
  wave speed.
\end{abstract}

\maketitle{}

The quest for a fundamental understanding of dynamic brittle fracture
has been an ongoing challenge.  The traditional continuum approach to
computing the speed of brittle cracks consists of solving the
equations of linear elasticity with boundary conditions on the moving
fracture surfaces up to the crack tip \cite{Freund}.  The solutions
have stress fields that diverge near the tip representing a finite
energy flow rate to the tip.  The crack speed $v$ is then assumed to
be uniquely determined by this energy flow rate.  In this theory, all
the nonlinear physics of failure inside a microscopic region around
the tip $-$ the so-called process zone $-$ is buried in a
phenomenological function that relates the fracture energy $\Gamma$ of
the material (energy needed to advance the tip per unit length of
crack front and per unit of crack extension) and $v$.

Detailed experiments have been carried out to test this theory by
measuring the speed of accelerating cracks with different initial
lengths and different loads \cite{exp3}.  The theory should be valid
if a unique $\Gamma(v)$ curve applies to the different cracks for the
same material.  The main difficulty in any experimental test of this
theory is that the energy flux to the tip cannot be measured directly.
Therefore, this flux needs to be inferred from a time-dependent
solution of linear elasticity for accelerating cracks.  Using an
approximate solution, Sharon and Fineberg have concluded that the data
for different cracks collapse on a unique $\Gamma(v)$ curve, whereas,
using an exact solution of Eshelby \cite{Esh69}, Kessler and Levine
have concluded that this collapse does not occur \cite{KesLev02}.

Also of long-standing interest in brittle fracture is the existence of
a dynamic instability that limits the speed of fast moving cracks.
This generic instability has been seen in both experiments
\cite{exp3,cryst,exp1} and molecular dynamic simulations \cite{md}.
The fact that different amorphous materials with entirely different
microscopic details (such as glass and PMMA) exhibit strikingly
similar branching instabilities \cite{exp3} strongly suggests that a
continuum theory may be appropriate for understanding this phenomenon.
Devising such a theory, however, has proven to be difficult.  Cohesive
zone theories modify the boundary conditions on the stress field near
the tip to take into account the short scale force between crack
surfaces.  Langer and Lobkovsky \cite{LanLob98}, however, have shown
that these models are unsuitable for stability calculations since the
results depend singularly on the details of the cohesive zone.  In
addition, intrinsically discrete branching instabilities in lattice
models \cite{MarLiu93,KesLev01} seem unlikely to bear relevance to
experiments in amorphous materials.

In this letter, we study dynamic brittle fracture using a recently
developed continuum phase-field approach for mode III cracks
\cite{Karetal01}.  This approach has the chief advantage that it
incorporates both the short-scale physics of failure and macroscopic
linear elasticity within a self-consistent set of coupled nonlinear
partial differential equations that can be solved numerically.
Moreover, solutions free of discretization artifacts can be obtained.
Earlier simulations of this model were limited to small system sizes
owing to the extreme stiffness of the equations, and no branching
instability was found even for cracks approaching the wave speed
\cite{Karetal01}.  Here, we briefly discuss how to overcome this
stiffness and carry out simulations in systems large enough to make
contact with results from the fracture community.  In addition, these
simulations enable us to study the onset of the characteristic
branching instability at a fraction of the wave speed that turns out
to be a robust feature of this phase-field model.
 
The basic variables of the model \cite{Karetal01} are the scalar
displacement $u(x,y)$ perpendicular to the $x$-$y$ plane of mass
points from their original positions, and the phase-field,
$\phi(x,y)$, which describes the state of the material.  The unbroken
solid, which behaves purely elastically, corresponds to $\phi = 1$,
whereas the fully broken material that cannot support stress
corresponds to $\phi = 0$.  The total energy (kinetic plus elastic) of
the system per unit length of the crack front is
\begin{equation}
  \label{eq:energy}
  E = \int dx \,dy \, \left[ \frac{\rho}{2}\left(\frac{\partial
        u}{\partial t}\right)^2 + \frac{\kappa}{2} \, |\vec \nabla
    \phi|^2 + h f(\phi) + \frac{\mu}{2}\, g(\phi) \left(|\vec \epsilon|^2 -
      \epsilon_c^2\right)\right],
\end{equation}
where $\rho$ is the density, $\vec \epsilon \equiv \vec \nabla u$ is
the strain, $f(\phi)$ is a double-well potential with minima at $\phi
= 1$ and $\phi = 0$, $\mu$ is the elastic shear modulus, and
$\epsilon_c$ is the critical strain magnitude such that the unbroken
(broken) state is energetically favored for $|\vec \epsilon| <
\epsilon_c$ ($|\vec \epsilon| > \epsilon_c$).  The function $g(\phi)$
is a monotonously increasing function of $\phi$ with limits $g(0) = 0$
and $g(0) = 1$, which controls the softening of the elastic energy as
more bonds are broken.

Taking the first variations of the energy with respect to the strain
and to $\phi$, we obtain the stress, $\vec \sigma=\delta E/\delta \vec
\epsilon$, and the equations of motion
\begin{subequations}
  \label{eq:pf}
  \begin{eqnarray}
    \frac{\partial \phi}{\partial t} & = & -\chi \, \frac{\delta
      E}{\delta  \phi},    
    \label{eq:pf1} 
    \\
    \rho \, \frac{\partial^2 u}{\partial t^2} & = & 
    \vec\nabla \cdot  \vec \sigma.
    \label{eq:pf2}
  \end{eqnarray}
\end{subequations}
Energy is dissipated in the process zone around the crack tip where
$\phi$ varies rapidly in space and time.  Eq.~(\ref{eq:pf1}) implies
that the size of the process zone is $\sim \xi =
\sqrt{\kappa/\mu\epsilon_c^2}$ and the characteristic time of energy
dissipation in this zone is $\tau = 1/(\chi \mu \epsilon_c^2)$.
Furthermore, by rescaling lengths by $\xi$, time by $\xi/c$, where $c
\equiv \sqrt{\mu/\rho}$ is the shear wave speed, and $u$ by $\xi
\epsilon_c$ in Eqs.~(\ref{eq:pf}), we find that crack propagation is
controlled by two dimensionless parameters $\delta = h/(\mu
\epsilon_c^2)$ and $ \beta = c \tau/\xi $.  The first determines the
surface energy normalized by $\mu \epsilon_c^2\xi$ \cite{Karetal01}
\begin{equation}
  \widetilde\gamma \equiv \frac{\gamma}{\mu \epsilon_c^2\xi}=  
  \int_0^1 d\phi\, \sqrt{1 -
    g(\phi) +  2 \delta f(\phi)}.
  \label{eq:gamma}
\end{equation}
For the choices $f(\phi) = 16 \phi^2 (1 - \phi)^2$ and $g(\phi) =
4\phi^3 - 3\phi^4$ here, $\widetilde\gamma$ approximately doubles when
$\delta$ increases from $0$ to $2$.

Parameter $\beta$ controls the importance of inertia relative to
dissipation in the process zone.  When $\beta \ll 1$
(inertia-dominated regime), failure is rapid and crack propagation is
governed by elastic energy flow.  When $\beta \gg 1$
(dissipation-dominated regime), failure is sluggish.  The elastic
displacements are quasistatic and propagation is governed by the
kinetics of the failure process.
  
We study fracture numerically in a strip of width $2W$ with a fixed
displacement $u(x,\pm W)=\pm \Delta$ at the strip edges.  The stored
elastic energy per unit area ahead of the crack tip is $G = \mu
\Delta^2/W$ and the Griffith's threshold load for a semi-infinite
crack is $G_c = 2\gamma$.

The equations of the model are inherently stiff due to strain
localization behind the crack tip \cite{Karetal01}.  The strain is
localized within a narrow region whose width vanishes in the large
strip limit $W/\xi\rightarrow \infty$, while the width of the $\phi$
variation remains of order $\xi$ in this limit.  Therefore, while it
is possible to resolve the spatial variation of $\phi$ everywhere,
resolving the strain concentration behind the crack tip is
computationally impractical for large strip widths.  The key to
overcoming this difficulty is to recognize that the model remains
well-defined when the strain variation is resolved only in the process
zone, but is allowed to be discontinuous on the lattice scale far
behind the tip.  Therefore, it is possible to simply discretize the
energy $E$ on a square lattice and to formulate the dynamics of the
model from partial derivatives of $E$ with respect to the variables
$\phi$ and $u$ at lattice points.  One subtlety of the present
simulations is that high-frequency waves are radiated behind the crack
tip above a small threshold tip speed.  These waves increase slightly
the fracture toughness and originate from the region behind the tip
where the displacement becomes discontinuous on the lattice scale. In
addition, multiple reflections of these waves can cause tip
oscillations at long times.  This time, however, is longer than the
time for propagation to reach a steady-state or branching to occur
such that this feature is not problematic.  Furthermore, this
radiation is easily suppressed by adding a small Kelvin viscosity to
Eq.~(\ref{eq:pf2}) with no qualitative change of the results
presented here.

\begin{figure}[t]
  \centering
  \includegraphics[width=12cm]{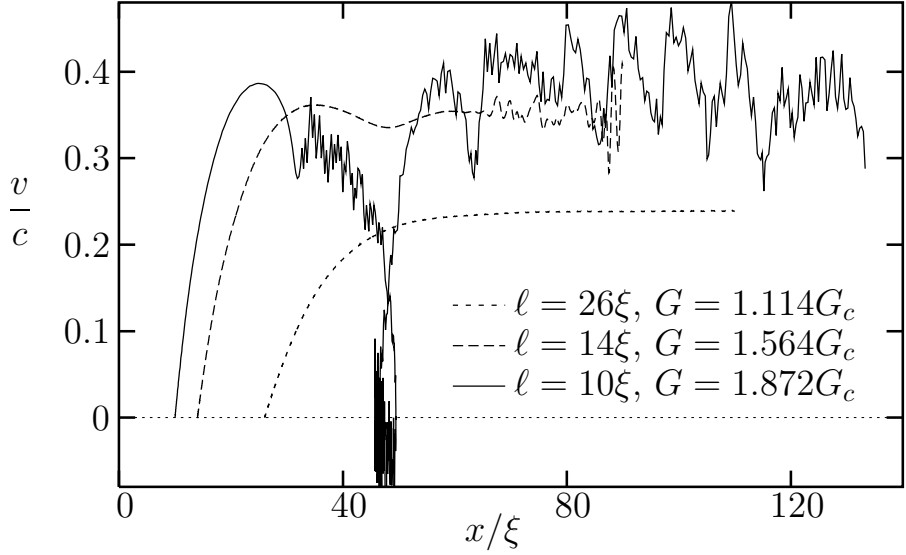}
  \caption{Crack velocity vs.~tip position for cracks accelerating
    from rest with different initial lengths.  $\beta = 1$, $\delta =
    0$, lattice spacing $dx = 0.3\xi$, and $W = 30\xi$.  The velocity
    for the largest load becomes negative since the symmetric branches
    retreat to give way to the asymmetric tip-splitting mode with a
    sinusoidal (snake-like) fracture path.}
  \label{fig:v_x}
\end{figure}

We have verified that the important observables converge quickly in
the limit of the vanishing lattice spacing, and that we operate in the
regime where these observables differ by at most 15\% from their
continuum limits for the lattice spacing $dx$ in the range $0.3\xi$ to
$0.4\xi$.

To study unsteady crack propagation, we use stationary solutions of
Eqs.~(\ref{eq:pf}) as initial conditions that correspond to cracks of
different initial length at rest.  For a given load $G$ and
corresponding initial crack length $\ell$, these stationary solutions
are found by relaxing $u$ with the Gauss-Seidel iteration scheme,
which amounts to solving Eq.~(\ref{eq:pf2}) without inertia, while
$\phi$ is relaxed using Eq.~(\ref{eq:pf1}).  This procedure yields
stationary cracks whose initial length $\ell$ decreases with
increasing $G$.  We then simulate the full equations of motion with
inertia to study accelerating cracks with zero initial velocity.
These simulations parallels previous experiments \cite{exp3} and
lattice simulations \cite{KesLev02}.  To study steady-state features
of crack propagation independent of initial conditions, we run long
simulations in strips that are effectively infinite along the
propagation direction.  To keep the computations tractable, we
periodically translate the fields $u$ and $\phi$ by one lattice
spacing such that the crack tip remains in the middle of a strip of
length much larger than $W$.  We have checked that the results of
these ``tread mill'' simulations are independent of boundary effects.

To compute the fracture energy during unsteady motion, we equate
$v\Gamma$ with the expression for the energy flow rate to the tip of
mode III cracks \cite{Esh69}
\begin{equation}
  v\Gamma = \int_C dC \left[\mu \,\dot u\,\partial_n u + v \, n_x \,
    \partial_t \left(\rho\dot u^2/2+\mu|\vec \epsilon|^2/2
    \right)\right],
\end{equation} 
where $C$ is a closed circuit around the moving tip and $\hat n$ is
the outward normal to $C$.  Since energy is conserved in the model
where $\phi = 1$, we can equivalently obtain the energy flow rate to
the tip by calculating the time rate of change of the total energy
energy (elastic plus kinetic) in the region of the system where $\phi$
is larger than a threshold value $\phi_c$ arbitrarily close to unity.
This quantity is precisely the energy flow rate into the process zone
defined as the region where $\phi < \phi_c$.

\begin{figure}[t]
  \centering
  \includegraphics[width=12cm]{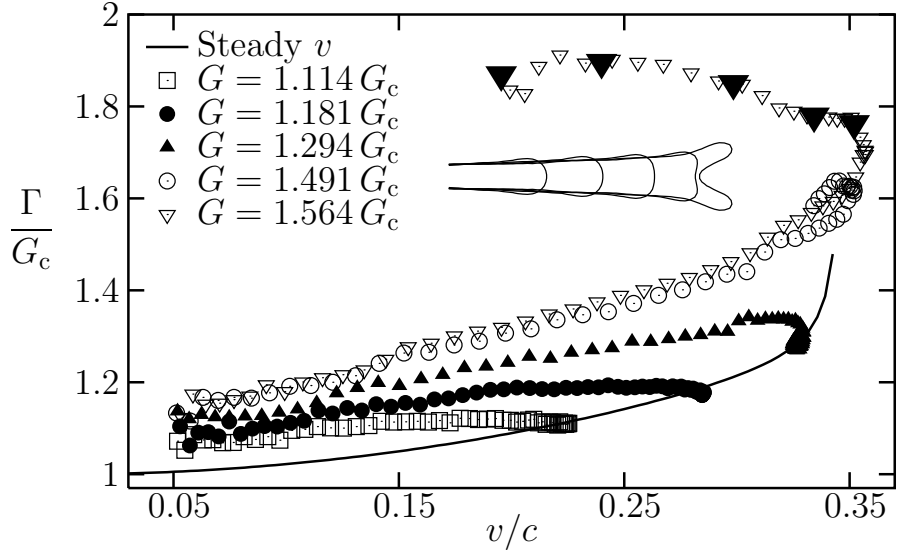}
  \caption{Fracture energy vs.~velocity for the 
    accelerating cracks and the same parameters as in Fig.
    \protect\ref{fig:v_x}. The solid line corresponds to steady-state
    crack propagation with $\Gamma=G$ by energy conservation.  The
    five $\phi = 1/2$ contours shown on the inset correspond to the
    five large solid triangles.  They are equidistant in time
    separated by $10\xi/c$.  Blunting and subsequent splitting is
    responsible for the deceleration.}
  \label{fig:GV}
\end{figure}

Plots of tip speed versus tip position for cracks accelerating from
rest are shown in Fig.~\ref{fig:v_x} for the inertia-dominated regime
$\beta = 1$.  Corresponding plots of $\Gamma$ versus $v$ are shown in
Fig.~\ref{fig:GV}.  The initial crack acceleration increases with load
and the crack tip splits into two symmetric branches above a critical
onset load $G_\mathrm{onset}$.  Cracks for smaller loads which do not
split, or split only transiently for $G\sim G_\mathrm{onset}$, reach a
steady-state propagation velocity that coincides with the steady-state
velocity calculated on the tread mill (solid line in
Fig.~\ref{fig:GV}).  The fact that $\Gamma=G$ in steady-state is a
self-consistency check of our method of calculating $\Gamma$ for
unsteady cracks.

\begin{figure}[t]
  \centering
  \includegraphics[width=12cm]{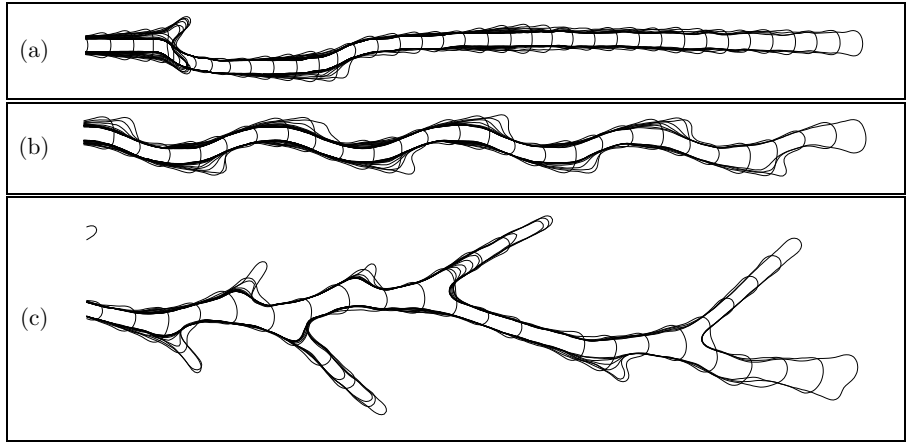}
  \caption{Contours of $\phi = 1/2$ separated in time by $\Delta t =
    10 \xi/c$.  All plots are drawn on the same scale for the
    parameters $\beta = 2$, $\delta = 0$, and $W = 30\xi$. (a)
    Transient branching for $G = 1.56 G_\mathrm{c}$.  (b) Weak
    periodic branching (the ``snake'') for $G = 1.86 G_\mathrm{c}$.
    (c) Chaotic branching for $G = 2.90 G_\mathrm{c}$.}
  \label{fig:snapshots}
\end{figure}

From the long simulations on the tread-mill, three basic regimes of
crack propagation can be distinguished: a stable regime for small
loads, where $v$ is constant in time and the crack is rectilinear, an
asymmetric tip-splitting (``snake'') mode for intermediate loads,
where $v$ fluctuates in time around some average value while the crack
follows a sinusoidal trajectory, and a chaotic tip-splitting regime
for large loads.  Examples of these regimes are shown in
Fig.~\ref{fig:snapshots}.  The simulations also reveal a strong
coupling between the crack tip dynamics and elastic waves that we have
not explored in detail.  A clear signature of this coupling is the
fact that the temporal period of branching in the snake mode is very
close to the period $4W/c$ of the first harmonic standing wave of the
strip.  The transient coupling of the tip to higher frequency waves is
also seen for higher loads in the chaotic branching regime.

\begin{figure}[t]
  \centering
  \includegraphics[width=12cm]{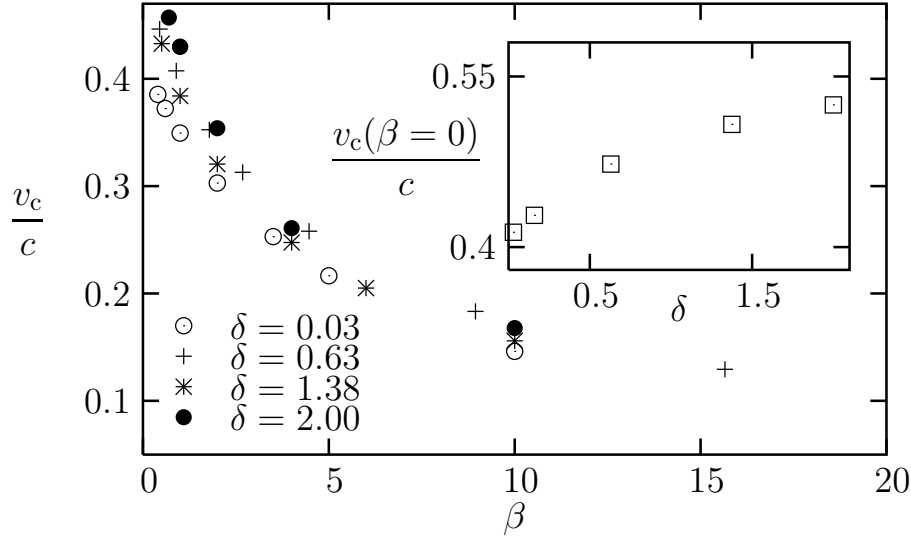}
  \caption{Limiting speed of crack propagation $v_c$ vs. $\beta$
    for different values of $\delta$ (or $\widetilde\gamma$ given by
    Eq.  \protect\ref{eq:gamma}).  Inset: $v_c$ in the
    inertia-dominated ($\beta \rightarrow 0$) limit vs.~$\delta$.}
  \label{fig:v_c}
\end{figure}

The steady state velocity $v$ saturates at some value $v_c$.
Rectilinear cracks cannot propagate faster than $v_c$.  We conclude
that off-axis branching in the present model is due to the absence of
steady-state crack solutions above a critical speed, rather than a
linear instability of solutions that exist up to the wave speed, as in
the lattice models \cite{KesLev01}.  The dependence of $v_c$ on the
parameters of the model is shown in Fig.~\ref{fig:v_c}.  The maximum
crack speed is well defined in the inertia dominated limit $\beta
\rightarrow 0$.  It grows monotonically with the scaled surface energy
$\widetilde \gamma$ (Eq.~(\ref{eq:gamma})) and has a minimum at
$\delta = 0$ of $v_c \approx 0.41 c$.  At this speed the linear
elastic field in the unbroken material around the crack tip is
quasi-isotropic \cite{Yoffe}.  We therefore conjecture, along the
lines of Gao \cite{Gao96}, that tip blunting which leads to velocity
saturation and ultimately to tip-splitting is due to the relativistic
contraction of stress fields in the nonlinear process zone where the
sound speed is small due to the softening of the material.

The smallest value of $v_c \approx 0.41 c$ in the $\beta\rightarrow 0$
limit is consistent with the calculation of Adda Bedia~\cite{Add04}
which shows that tip splitting is energetically possible for mode III
fracture for speeds above $0.39 c$ with a maximum angle between
symmetric branches of 80 degrees, which is about 10-15 percent larger
than the angle in our simulations.  It should be noted, however, that
this calculation \cite{Add04} considers a steady-state crack that
stops abruptly and splits into two daughter cracks, whereas cracks
decelerate before splitting in our simulations.  In addition, the
onset of branching depends here on the short-scale parameters $\beta$
and $\widetilde \gamma$.  Therefore, it cannot be predicted from
purely energetic considerations.

To interpret our results for accelerating cracks, it is useful to
first derive an analytic expression for the speed of steady-state
cracks close to the Griffith threshold.  Energy balance for
steady-state propagation implies that the stored elastic energy ahead
of the crack in excess of twice the surface energy must be dissipated
in the process zone.  This implies that
\begin{equation}
  v(G - G_c) = -\frac{dE}{dt} = \chi^{-1} \int d\vec x
  \left(\frac{\partial \phi}{\partial t}\right)^2,
  \label{eq:Edot}
\end{equation}
where the second equality follows from Eqs.~(\ref{eq:pf}) with no
energy flux through the boundaries.  In a coordinate system ($x' = x -
vt, \, \, y' = y$) translating with the tip, $\partial \phi/\partial
t= -v\, \partial \phi/\partial x'$.  Thus the steady-state velocity of
rectilinear crack propagation is
\begin{equation}
  \frac{v}{c} \approx \frac{2\tilde \gamma (G/G_c - 1)}{\beta \, I},
  \label{eq:v_just_above}
\end{equation}
where $I\equiv \int dx'dy' \left(\partial \phi/\partial x'\right)^2 $
is a dimensionless integral factor of order unity that depends on the
profile of $\phi$ for a stationary crack at $G\approx G_c$. This
prediction is consistent with the simulation results that show a
steeper increase of velocity with load for smaller $\beta$.

The fact that the $\Gamma-v$ plots for different accelerating cracks
in Fig.~\ref{fig:GV} do not superimpose on the steady-state curve
clearly shows that the speed of these cracks is not uniquely
determined by the energy flow rate to the tip.  In our model,
accelerating cracks require more energy per unit length of advance
than steadily propagating cracks.  This extra energy is reversibly
stored in the process zone to be either consumed by branching or
radiated away later.  This effect is exacerbated for large loads and
causes the crack to decelerate before branching.  This deceleration is
marked by the overshoot of the velocity plotted as a function of tip
position (Fig.~\ref{fig:v_x}) and the fact that the corresponding
$\Gamma-v$ plot becomes double-valued (Fig.~\ref{fig:GV}).  A
qualitatively similar overshoot of the velocity was observed
experimentally for very large accelerations in brittle fracture of
glass \cite{exp3}, albeit for much larger systems than in the present
simulations.

To estimate when crack acceleration should be important in our model,
we compute the change of the velocity over a distance comparable to
the process zone.  Crack's acceleration should be irrelevant when this
change is small compared to the wave speed, or $\xi \, d(v/c)/dx \ll
1$.  For cracks accelerating from rest, let us assume that
Eq.~(\ref{eq:v_just_above}) gives the instantaneous speed of the
accelerating crack if $G$ is taken to mean the instantaneous elastic
energy release rate $\Gamma$ which varies as $d\Gamma/dx \sim (G -
G_c)/W$ as the crack extends into the stressed strip.  This implies
that the acceleration is small when
\begin{equation}
  \frac{W}{\xi} \gg \frac{G - G_c}{G_c\beta}.
  \label{eq:small_acc}
\end{equation}
We find that acceleration remains important in our simulations even
for $W/\xi=100$ when $\beta$ is of order unity.  Since in experiments
$G$ can be up to a hundred times larger than $G_c$ and
$\beta=c\tau/\xi$ can be much smaller than unity, acceleration may be
important even in macroscopic strips several orders of magnitude
larger than the process zone size.  We conclude that the validity of
the continuum theory of brittle fracture depends not only on the ratio
of the system size to the process zone size, but also on the
importance of inertia relative to bond breaking dissipation measured
here by $\beta$.

This conclusion does not conflict with the fact that this theory was
recently validated by lattice simulations in the limit of zero
dissipation \cite{KesLev02}.  In these simulations, the velocity jumps
discontinuously to a fraction of the wave speed and the slope of the
velocity-load relation is finite for velocities larger than this
fraction.  Therefore, $1/\beta$ should be replaced by
$d(v/c)/d(G/G_c)\sim 1$ in the above estimate leading to
Eq.~(\ref{eq:small_acc}).  In contrast, for the velocity range from
zero to a fraction of the wave speed pertinent to experiments, the
continuum theory will break down for accelerating cracks in the
$\beta\rightarrow 0$ limit of the present model for any finite system
size.  The crack dynamics is independent of $\beta$ and well-defined
in this limit.

Clearly, the incorporation of more realistic microscopic mechanisms of
failure in the present phase-field approach and the extension to mode
I remain needed to make contact quantitatively with experiments.

This research is supported by U.S. DOE Grant No. DE-FG02-92ER45471.

\end{document}